\documentstyle[twocolumn,aps,prb,psfig]{revtex}
\begin{document}
\draft
\title{
Ising Dipoles on the Triangular Lattice
}
\author {\it U.K.\ R{\"o}{\ss}ler
}
\address{
IFW Dresden, Institut f{\"u}r Metallische Werkstoffe,\\
Helmholtzstra{\ss}e~20, D-01069 Dresden, Germany.
}
\date{October 10, 2000}
\maketitle
\begin{abstract}
A cluster Monte Carlo method for systems of classical spins 
with purely dipolar couplings is presented. 
It is tested and
applied for finite arrays of perpendicular Ising dipoles 
on the triangular lattice. This model is a modification 
with long-range interactions of the geometrically 
frustrated Ising antiferromagnet.
From measurements of integrated autocorrelation times for energy,
magnetization and staggered magnetizations, a high efficiency 
of the cluster Monte Carlo (MC) algorithm compared 
to a single-spin-flip algorithm is found.
For the investigated model, a finite temperature
transition is found which is characterized by a peak
in the specific heat and in the staggered susceptibilities.
\end{abstract}

The progress in the fabrication of very small patterns 
of ferromagnetic materials
\cite{Kent 93 94,%
Scheinfein 96,%
Wirth 98 99 00a,%
Cowburn 99}
opens the fascinating possibility 
to study experimentally statistical properties 
of some mesoscopic or exotic magnetic systems.
Dipolar couplings between such particles may induce 
magnetic ordering \cite{Wirth 98 99 00a,Cowburn 99}.
These couplings are often only viewed as 
additional effective anisotropy barriers 
for the superparamagnetic relaxation 
of individual particles.
However, such assemblies of ferromagnetic
nano-particles are coupled magnetic systems.
They should display magnetic phase-transitions 
for systems with mesoscopic 
size of individual magnetic moments.
Recently, Wirth et al. studied the properties
of nanoscale ferromagnetic columns which
were grown on substrates 
in a scanning tunnel-microscope\cite{Wirth 98 99 00a}. 
Magnetic-force microscopic 
pictures of densely packed columns 
on a triangular lattice showed typical
antiferromagnetic correlations 
where rows of nearest-neighbour moments
point alternatingly up and down.

This paper presents a numerical study 
on the expected thermodynamics for such 
assemblies of perpendicular dipoles in
two dimensions. Its second aim is the
introduction of a cluster Monte Carlo (MC)
method for dipolarly coupled systems.
For the geometrically frustrated triangular
lattice, this method allows efficient statistical 
sampling down to relatively low temperatures.
The coupled system we want to study consists of 
$i=1\dots N$ Ising--spins, $\sigma_i \in \{-1,1\}$
\begin{equation}\label{IsingDipoles}
H=\frac{1}{2}\,{\sum_{i,j}}'\,
              v(R_{ij}) {\sigma}_i {\sigma}_j\,,
\end{equation}
where $v(R)=g\, R^{-3}$
are interactions 
between three--dimensional dipoles 
describing the coupling between
the perpendicular moments of 
the particles. 
All the couplings are antiferromagnetic.
Length is measured in nearest-neighbour
distances. If we restrict the interactions 
in (\ref{IsingDipoles}) to nearest-neighbour
terms we get an Ising antiferromagnet.
The triangular Ising antiferromagnet
(TIAFM) is one of the few exactly 
solved systems with geometrical frustration 
\cite{Wannier 50 Stevenson 70}.
It was shown that no long-range
order is established at finite temperature,
but the system becomes critical at $T=0$.
However, additional interactions 
may lift the ground-state degeneracy
and may induce phase-transitions in
such underconstrained systems.
Hence, model (\ref{IsingDipoles})
is interesting in its own right.

Cluster-MC methods for spin models
\cite{Swendsen Wang 87,Wolff 89,Kandel Domany 91}
are only efficient in critical regions 
if the clusters do not percolate already 
above the critical temperature.
The energy of model (\ref{IsingDipoles})
may be decomposed into two parts
\begin{equation}\label{ModelPartitions}
H=\sum_{\Delta} V_{\Delta} + 
\frac{1}{2}\,{\sum
\limits_{(i,j)\,\not\in\, {\rm N\!N}}}\!\!\!\!\!'\, v(R_{ij})\,,
\end{equation}
where the sum of $V_{\Delta}$ denotes 
sums of nearest-neighbour interactions
in one of the two types of triangular plaquettes 
(and, with open boundary conditions, some 
nearest-neighbour interactions along the boundaries
also labeled by  $V_{\Delta}$). 
This first part is exactly the TIAFM.
The second sum contains the remaining bonds with longer range.

Clusters for the TIAFM
can be defined 
by sampling the sums of energies
corresponding to elementary
triangular plaquettes
\cite{Coddington Han 94 Zhang Yang 94}.
These clusters are the correct critical clusters
of the TIAFM
percolating at $T=0$.
Applying this cluster definition to
the first part of (\ref{ModelPartitions}) 
the dipolar system is mapped on a system
of clusters which interact through the 
long-range interactions
generated from terms in the second sum of (\ref{ModelPartitions}).
For flipping clusters 
the usual Metropolis algorithm is used. 
The algorithm of alternatingly mapping (1) 
onto interacting clusters via (2) 
and sampling this mapped system 
is equivalent to sampling the original system.

We have to show ergodicity
and detailed balance to justify the method. 
The procedure of cluster decomposition may result 
in deleting all bonds which maps the original system 
onto itself. Thus, the occurrence of single-spin flips 
ensuring ergodicity can be checked {\em a posteriori}.
Detailed balance may be demonstrated 
by the prescription for general cluster algorithms 
given by Kandel and Domany\cite{%
Kandel Domany 91} which shows 
that bonds may be deleted, frozen,
or left unchanged to generate 
systems of interacting clusters 
equivalent to the original system.

The method is applied 
for finite arrays 
of hexagonal shape 
with $N=61 \dots 2977$ spins.
The following measure for 
antiferromagnetic correlations
is used:
Spins are labeled $\sigma_{\mu\nu}$
for sites given 
by $\mu\,{\bf r}_1+\nu\,{\bf r}_2$
with nearest-neighbour vectors ${\bf r}_{1,2}=(\pm\sqrt{3}/2,1/2)$.
Three types of staggered magnetizations
may be calculated
by 
$m^{\mbox{\scriptsize I}}=%
1/N\,\sum_{\mu\nu}\,\sigma_{\mu\nu}(-1)^{\mu}$,
$m^{\mbox{\scriptsize II}}=%
1/N\,\sum_{\mu\nu}\,\sigma_{\mu\nu}(-1)^{\nu}$,
and 
$m^{\mbox{\scriptsize III}}=%
1/N\,\sum_{\mu\nu}\,\sigma_{\mu\nu}(-1)^{\mu+\nu}$.
For the hexagonal arrays, the three types are equivalent.
Thus we use the average 
$m_a=%
1/3\;(m^{\mbox{\scriptsize I}}+%
m^{\mbox{\scriptsize II}}+%
m^{\mbox{\scriptsize III}})$.

Here, our algorithm is first used
in the Swendsen-Wang (SW) form 
by deleting/freezing bonds for 
all plaquettes of a randomly chosen type. 
It turns out that after mapping onto the system
of interacting clusters best performance
was achieved by attempts to flip just
one (randomly chosen or the largest) cluster.
Integrated auto-correlation times for energy, 
magnetization, 
and for the staggered magnetizations 
$m^{\mbox{\scriptsize I}\,\dots\,\mbox{\scriptsize III}}$
were calculated 
as measure of efficiency 
by a self-consistent 
windowing method \cite{Wolff 89a}.
Independent isothermal runs were started
from slowly cooled and relaxed configurations.
For sizes $N=61 \dots 469$ between 
several $10^4$ and $10^5$ cluster-MC
sweeps (MCS, i.e., $N$ attempted cluster moves),
for the two larger sizes only $10^{3\,\dots\,4}$~MCS 
could be performed.
As expected, the autocorrelations 
for the magnetization decay rapidly 
at all temperatures.
The autocorrelations for energy are 
similar to those of the staggered magnetizations.
Thus, we only show the average integrated autocorrelation 
time $\tau_a$ for the staggered magnetizations 
in Fig.~1(a) and compare it to runs 
with the single-spin-flip Metropolis algorithm for 
the two smallest system sizes.
Fig.~1(b) shows the average 
size $<\!c\!>$ of flipped clusters. 
The strong increase of $<\!c\!>$ towards low temperature
means that all spins are increasingly correlated. 
This transition becomes sharper with larger $N$.
The size distribution of flipped clusters always consists of
a rapidly decreasing part starting with single-spin flips.
For low temperatures and small system sizes clusters 
with size close to $N$ are found also.

Still, the corresponding $\tau_a$ does not 
grow dramatically. For $N>469$ we could 
measure $\tau_a$ only down to $T/g \simeq 0.19$, 
where it becomes essentially 
independent of $N$ while $<\!c\!> \simeq 2$.
The single-spin-flip algorithm yields
far larger $\tau_a$  which in a critical 
region should grow with system size as
$\tau_a \propto N^{z/2}$ with $z\simeq 2$.
This comparison demonstrates the efficiency 
of this cluster-MC.
However, the numerical cost of the SW-algorithm 
scales as $N\,\tau_a$.
Therefore, a Wolff-type algorithm \cite{Wolff 89}
was implemented for further production 
runs shown below.
Only a single cluster is grown starting 
from a randomly chosen plaquette.
The performance of this algorithm is 
similar to the SW-algorithm but
the numerical effort is smaller. 
Its numerical cost scales as $<\!c\!>\,\tau_a$.

Cluster-MC runs with several $10^5$~MCS 
were performed for $N\le 721$.
For $N\ge 1519$ serious problems 
in the critical region $T/g \le 0.18$ 
arise. Cycling temperature displays 
hysteresis of the energy.
Measurements could be done for runs 
with only few $10^3$~MCS 
in the critical region with $N\ge 1519$.
Here, configurations both from cooling 
and heating were first relaxed for longer times
before start of measurements to ensure that
they are close to equilibrium.
Sampling was performed
for energy per spin $E$,
specific heat $C=N(<\!E^2\!> - <\!E\!>^2)/T^2$,
the staggered magnetizations,
the corresponding susceptibilities
$\chi^{\mbox{\scriptsize X}}%
=N(<\!(m^{\mbox{\scriptsize X}})^2\!>-<\!m^{\mbox{\scriptsize X}}\!>^2)/T$
and the fourth-order Binder cumulant
$U^{\mbox{\scriptsize X}}=%
1-<\!(m^{\mbox{\scriptsize X}})^4\!>/(3\,<\!(m^{\mbox{\scriptsize X}})^2\!>^2)$
(X $=$ I, II, III).
Using the histogram method 
by Ferrenberg and Swendsen \cite{Ferrenberg Swendsen 89}
these various quantities were evaluated
as functions of temperature 
in the critical region.
Results are shown in Figs.~2 and 3.
The diverging peak in the specific heat (Fig.~2(b)) 
indicates the presence of a transition
at or closely below $T/g \simeq 0.18$.
Similarly, the peak for the average staggered
susceptibility $\chi_a$ 
shows that the transition (Fig.~3(a)) 
is related to antiferromagnetic ordering.
Near  $T/g \simeq 0.18$, $m_a$ steeply 
increases reaching values $\sim$~0.3 to 0.4 (not shown here).
However, the average Binder-cumulant $U_a$ 
for the various system
sizes does not provide a clear indication
for a continuous transition by a unique
crossing of $U_a(T)$-curves 
for the different system sizes
at one critical temperature (Fig.~3(b)). 
The transition might be first order instead
as indicated by the observed hysteresis 
when cycling temperature. 
Then the apparent correlation of system 
sizes $N < 1000$ corresponds only
to the size of typical nuclei of the low-temperature phase.
On the other hand, the behaviour of the $U_a(T)$--curves
which seem to merge below $T/g=0.18$, 
is consistent with usual observations 
at a Berezinskii-Kosterlitz-Thouless-transition
and a line of critical points at low temperature.
Possibly, the peak in the specific heat 
is only a precursor of this transition,
or two transitions are present. 

In conclusion a cluster-MC method is presented
which may be used for dipolarly coupled spin-systems.
It works by generating interacting clusters 
from sampling of short-range interactions.
It is shown to be efficient for the specific
case of Ising dipoles on the triangular lattice. 
This success relies on the ability to define clusters 
for this geometrically frustrated system
which percolate only at $T=0$.
However, the method can
be generalized to various other systems with
long-range interactions, including systems with
continuous spins\cite{Wolff 89}.
The investigation of the exact nature 
of the finite-temperature transition found
with this MC-method remains a task for the future.

{\em Acknowledgment.} I thank S.\ Wirth and A.\ M{\"o}bius 
for discussion. This work was supported by DFG.
After finishing work on this paper 
I became aware that den Hertog et al.
have developed a similar cluster MC-method 
for dipolar spin ice 
on the pyrochlore lattice \cite{den Hertog 00}.
%


\begin{figure}[tb]
\psfig{%
figure=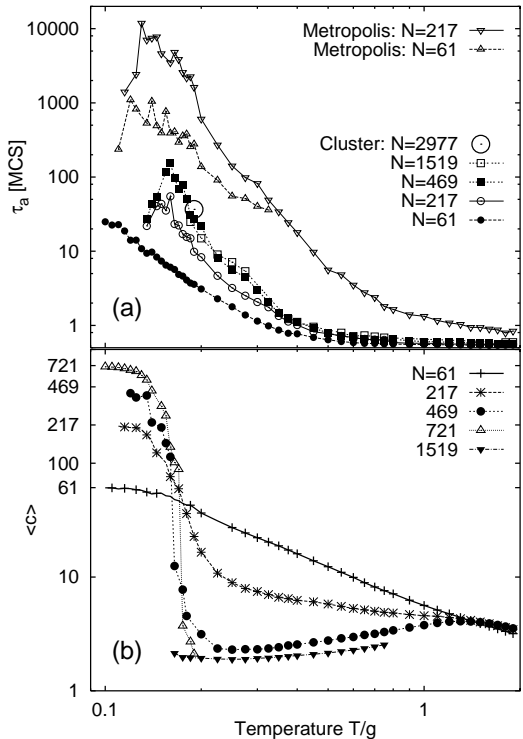,%
width=7.6cm}
\vspace{12pt}
\caption{(a) Average integrated autocorrelation time 
for staggered magnetizations (see text)
versus temperature. Symbols correspond to independent
isothermal runs. 
Comparison between Metropolis single-spin-flip 
and cluster algorithm. For the largest size $N=2977$, 
only one run at $T/g=0.190$ is shown (big circle).
(b) Corresponding average cluster sizes $<\!c\!>$ 
for different system sizes.}
\end{figure}
\begin{figure}[bt]
\psfig{figure=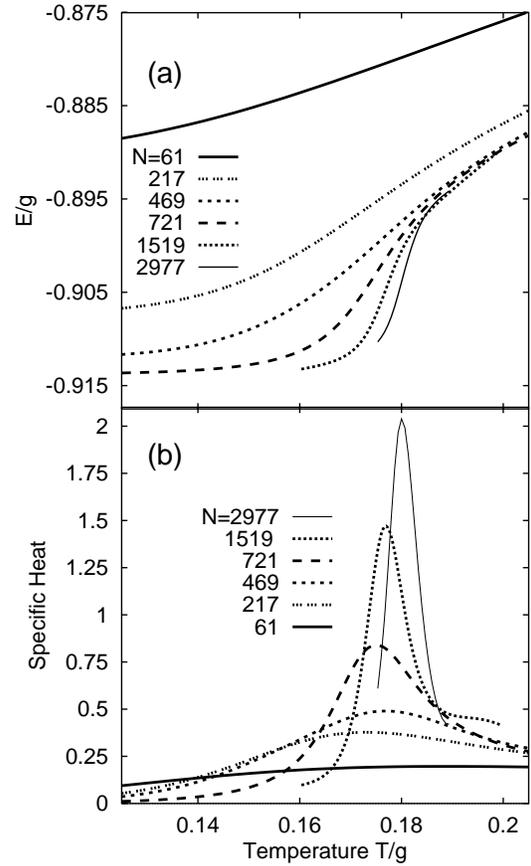,width=7.6cm}
\vspace{12pt}
\caption{(a) Energy per spin (b) specific heat
in the region of the transition temperature
of Ising dipoles on hexagonal triangular--lattice 
arrays with various sizes $N$.}
\end{figure}
\begin{figure}[tb]
\psfig{figure=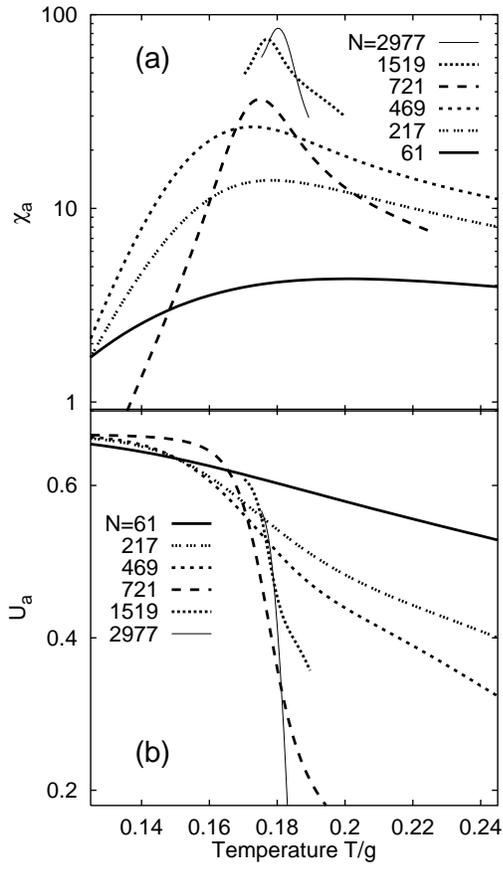,width=7.6cm}
\vspace{12pt}
\caption{(a) Average stag\-gered sus\-cep\-tibility
\(\chi_a=\) \(1/3\;(\chi^{\mbox{I}}\) \(+\chi^{\mbox{II}}\)
\(+\chi^{\mbox{II}})\).
(b) Corresponding average fourth-order Binder cumulant 
in the critical temperature region.}
\end{figure}

\end{document}